\documentclass[3p]{elsarticle}

\pdfoutput=1
\usepackage[unicode=true, bookmarks=false,bookmarksnumbered=true,bookmarksopen=true, breaklinks=false,backref=false,pagebackref=false, colorlinks=true]{hyperref}
\hypersetup{pdftitle={ },pdfauthor={ },linkcolor=black,citecolor=black, urlcolor=black, filecolor=black,pdfpagelayout=OneColumn,pdfnewwindow=true,pdfstartview=XYZ, plainpages=false}
\usepackage{color}

\usepackage{braket}
\usepackage[T1]{fontenc}
\usepackage{amsmath}
\usepackage{amsthm}
\usepackage{amssymb}
\usepackage{url}
\usepackage[english]{babel}
\usepackage{dcolumn}
\usepackage{bm}
\usepackage{subfig}
\usepackage{float}
\usepackage{url} 

\newcommand{\Tr}{\mbox{\rm Tr\,}}
\newcommand{\ReC}{\mbox{\rm Re\,}}

\newcommand{\be}{\begin{equation}}
\newcommand{\ee}{\end{equation}}
\newcommand{\bea}{\begin{eqnarray}}
\newcommand{\eea}{\end{eqnarray}}
\newcommand{\non}{\nonumber}
\newcommand{\bit}{\begin{itemize}}
\newcommand{\eit}{\end{itemize}}
\newcommand{\mbf}{\mathbf}

\graphicspath{{figures/}}
\usepackage{verbatim}

\begin{document}
\begin{frontmatter}

\title{Colour field flux tubes and Casimir scaling for various SU(3) representations}
\author{N. Cardoso}
\ead{nunocardoso@cftp.ist.utl.pt}
\author{M. Cardoso}
\ead{mjdcc@cftp.ist.utl.pt}
\author{P. Bicudo\corref{cor1}}
\ead{bicudo@ist.utl.pt}
\address{CFTP, Departamento de F\'{\i}sica, Instituto Superior T\'{e}cnico
(Universidade T\'{e}cnica de Lisboa),
Av. Rovisco Pais, 1049-001 Lisboa, Portugal}

\cortext[cor1]{Corresponding author}

\begin{abstract}

We investigate the QCD flux tubes linking static colour charges of different SU(3) representations, relevant to the understanding of confinement and of the Lund strings of Heavy Ion Collisions. 
The colour field densities, the Casimir scaling factors and the widths for the flux tubes of the first five different representations are computed in quenched SU(3) lattice QCD. This study is relevant to understand the mechanisms of confinement and also the flux tubes utilized in the Lund Model  and in other models of Heavy Ion Collisions.

\end{abstract}


\end{frontmatter}
\twocolumn

\section{Introduction}

We investigate the QCD flux tubes linking static colour charges of different SU(3) representations. 

Particle physics has been researching the building blocks of nature, now compiled in a single Standard Model, including the three fundamental quantum interactions.  These - the strong, electromagnetic and weak interactions - together with gravitation (so far we don't know exactly how to quantize it) can all be described by simple mathematical equations, using the Lagrangian technique. Both the Standard Model and the gravitation Lagrangians have quite precise analytical solutions in simple applications, say two-body problems or perturbative particle decays, except for the strong interaction. Because the internal symmetry group of the strong interaction is the non-commutative SU(3), the strong interaction Lagrangian is too non-linear to have any analytical solution. Nevertheless the strong interaction - also denominated Quantum Chromodynamics (QCD) - Lagrangian may have numerical solutions when we utilize the Lattice QCD techniques devised by Wilson
\cite{Wilson:1974sk}. 
Lattice QCD utilizes the Dirac-Feynman Path integral, the Wick rotation, and the Boltzmann statistical mechanics to transfer the QCD problems into an average over a succession of configurations, similar to parallel but extremely small classical universes.  Presently we are able to simulate lattices with sizes of the order of a few Fermi or femtometers. 

Confinement is one of the phenomena of QCD so far preventing its analytical solution. When a pair of QCD charge and anti-charge are slowly pulled apart, a flux tube of fields develops between the pair. Flux tubes, sometimes approximated by a thin string for modelling, are extended and non-linear objects, and they have been observed in Lattice QCD, both with SU(2) and SU(3) gauge groups
\cite{Fukugita:1983du,Flower:1985gs,Sommer:1987uz,Caldi:1988zj,DiGiacomo:1989yp,DiGiacomo:1990hc,Cardaci:2010tb,Michael:1985ne,Markum:1988na,Campbell:1985kp,Bali:1994de,Bissey:2009gw}.
Although these flux tubes are not directly observed experimentally, since we don't know how to create static enough flux tubes in the Laboratory, their evidence is found in the spectrum of mesons composed of a quark, an antiquark, and the flux tube these QCD charges produce.

\begin{table}[t!]
\begin{center}
\begin{tabular}{ccccc}
\hline
	$D$ & $(p,q)$ & $z^{p-q}$ & $p+q$ & $d_D$\\
\hline
\hline
	3 	& $(1,0)$	& $z$	& 1	& 1	\\
	8 	& $(1,1)$	& 1	& 2	& 2.25	\\
	6 	& $(2,0)$	& $z^*$	& 2	& 2.5	\\
	$15a$ 	& $(2,1)$	& $z$	& 3	& 4	\\
	10 	& $(3,0)$	& 1	& 3	& 4.5	\\
\hline
\end{tabular}
\end{center}
\caption{
Group theoretical factors for SU(3) as in \cite{Bali:2000un}. $D$ is the dimension
of the representation. $(p,q)$ are the respective weight factors and $z= \exp(2 \pi /3)$.
We compare our colour field ratios with the casimir scaling factor $d_D=C_D /C_3$, 
ratio of the quadratic Casimir charge of representation $D$ and the one of 
the fundamental representation 3.}
\label{tab:casimir}
\end{table}

\begin{figure*}[t!]
\begin{centering}
    \subfloat[ 3 ]{
    \includegraphics[width=5.0cm]{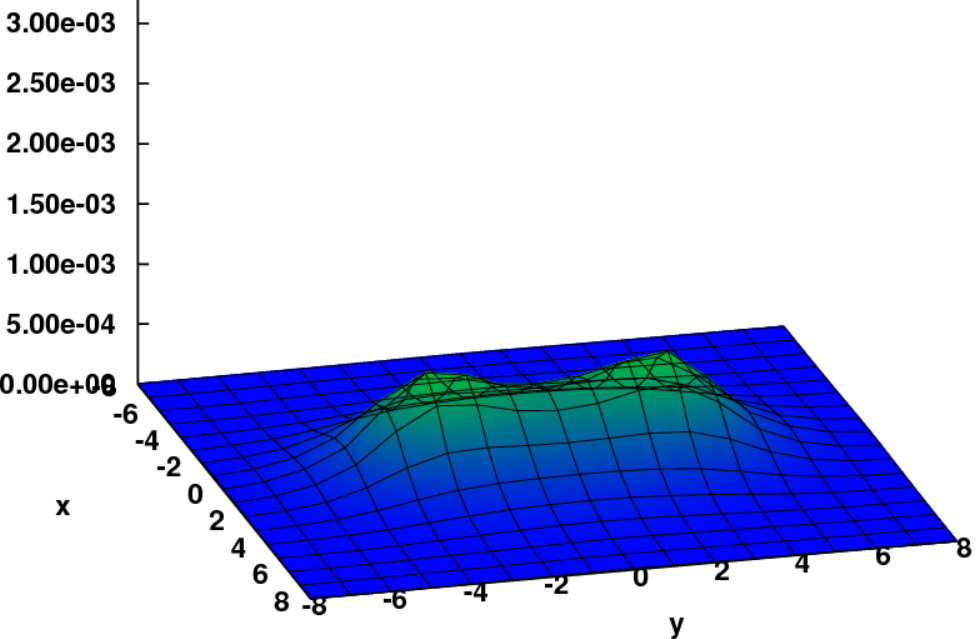}
    }
\hspace{.1cm}
    \subfloat[ 8 ]{
    \includegraphics[width=5.0cm]{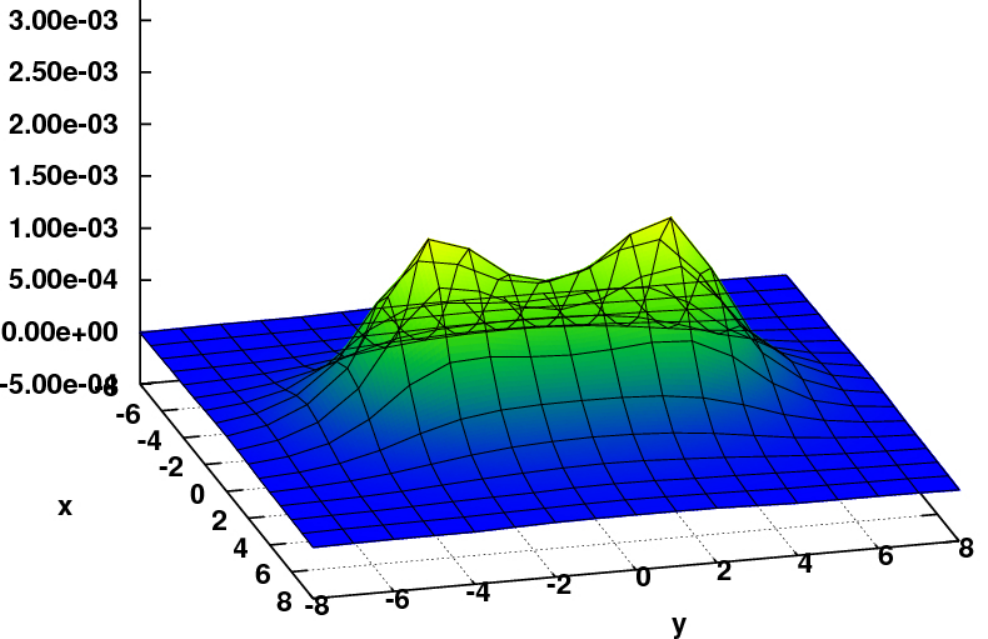}
     }
\hspace{.1cm}
    \subfloat[ 6 ]{
    \includegraphics[width=5.0cm]{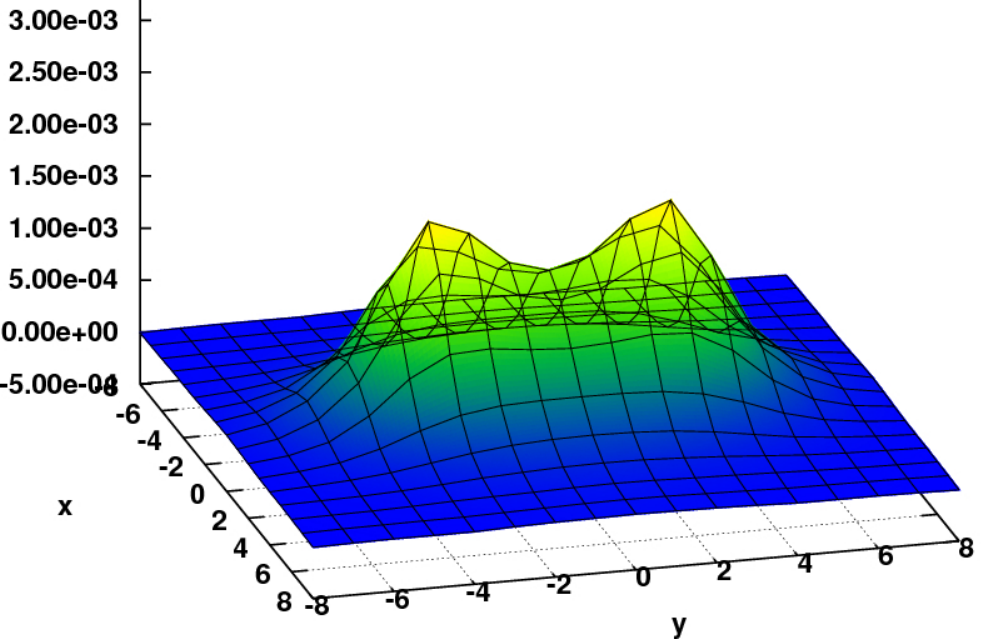}
    }
\end{centering}
\begin{centering}
    \subfloat[ 15a ]{
    \includegraphics[width=5.0cm]{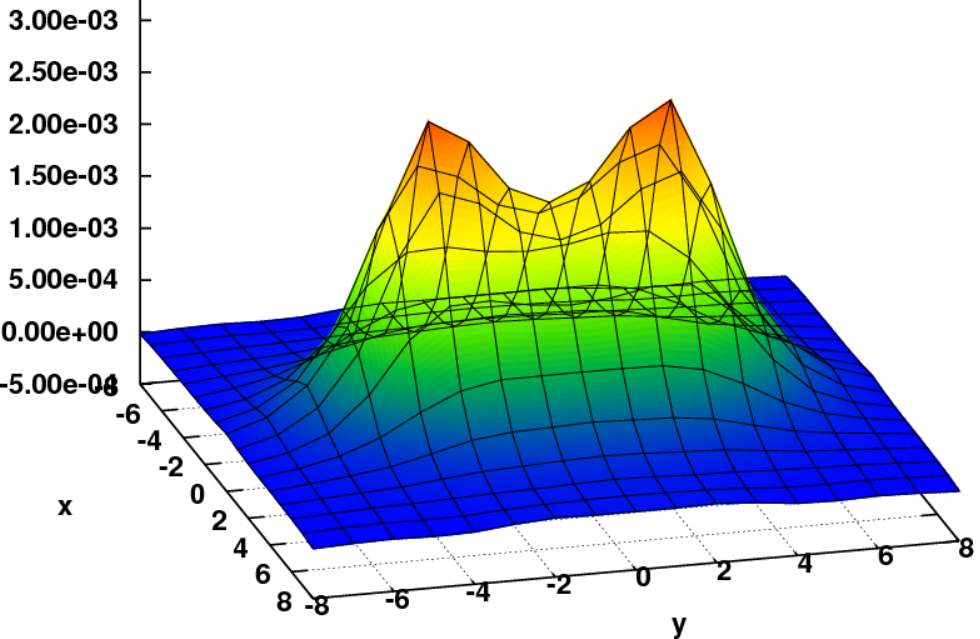}
     }
\hspace{.1cm}
    \subfloat[ 10 ]{
    \includegraphics[width=5.0cm]{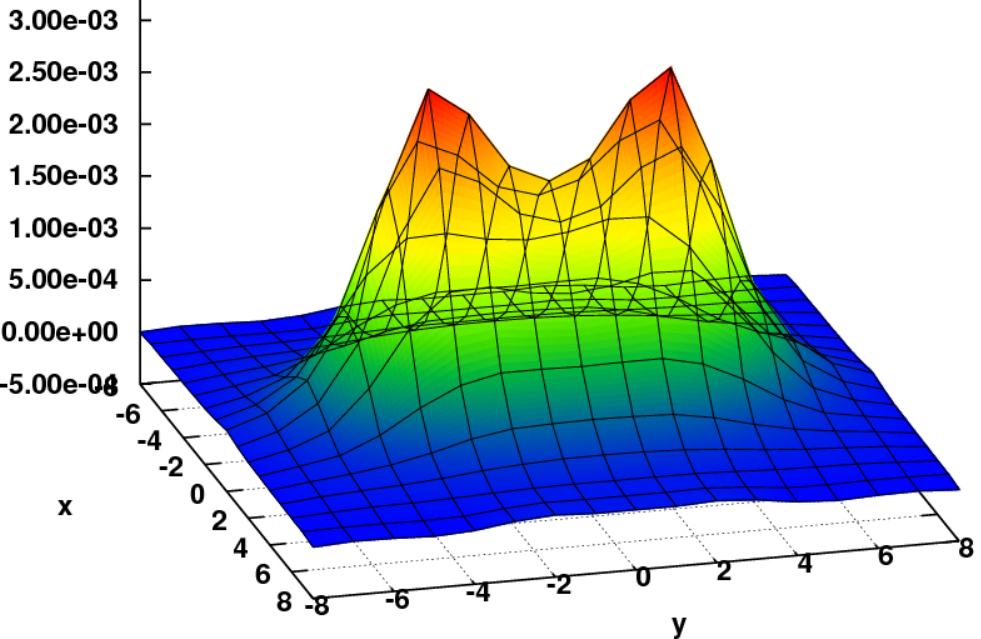}
    }
\hspace{2.1cm}
    \includegraphics[width=1.5cm]{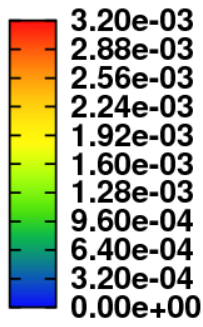}
\hspace{2.cm}
\end{centering}
    \caption{Lagrangian density created by static SU(3) charges of the representations, respectively from left to right:  3, 8, 6a, 15a, 10. The results are presented in lattice spacing units (colour online).}
    \label{fig:3D_Sim}
\end{figure*}

A new type of QCD flux tubes, different from the mesonic flux tubes, may also be indirectly observed in the Heavy Ion Collisions presently produced at the LHC and at the BNL. Strings or flux tubes are the building blocks of different models Heavy Ion Collision models. Heavy Ions, i. e. nuclei  composed of many particles, are collided at the highest possible energies, to approach in the Laboratory the extremely high temperatures occurring shortly after the Big Bang epoch in cosmological models. In the Lund hadronization model, 
\cite{Andersson:1983jt,Andersson:1983ia,Andersson:1998tv} the large number of particles produced in Heavy Ion Collisions are reproduced with string fragmentation.
Flux tubes also constitute the initial condition of recent hydrodynamic models describing the elliptic flow in Heavy Ion collisions
\cite{Mota:2011npa}. 
Surprisingly, the flux tubes utilized to simulate Heavy Ion Collisions may have a width as thin as 0.2 fm, and a string tension almost one order of magnitude larger than the fundamental string tension linking a mesonic quark-antiquark pair. 
Moreover, the Colour Glass Condensate
\cite{Iancu:2000hn,Iancu:2001ad},
produced at the onset of the Heavy Ion Collisions, is saturated by gluons. The quark degrees of freedom can be neglected at the onset of the quark gluon plasma, it is sufficient to study quenched SU(3) to address the Heavy Ion Collision flux tubes.
It has also been proposed that the gluonic partons saturating the Heavy Ion Collisions produce in the Colour Glass Condensate perturbative flux tubes with an original width, or transverse size, of the order of $1/ Q_s$, with $Q \simeq 1 \to 2$ GeV
\cite{Dumitru:2008wn,Dusling:2009ni,Lappi:2009vb}, 
flux tubes persisting during the evolution of the Quark Gluon Plasma.
Thus it is relevant to study the confining and non-perturbative flux tubes of higher SU(3) representations to understand if they are compatible with the flux tubes produced at the Heavy Ion Collisions.

The static potentials attracting a static colour charges and a colour anti-charge of different SU(3) representations have been already studied in quenched Lattice QCD. In the study of potentials, evidence was found for Casimir scaling in SU(2), first found by Ambjorn et al. 
\cite{Ambjorn:1984mb,Ambjorn:1984dp}.
Casimir scaling occurs when the string $\sigma$  of the linear component
of the static potential measured at intermediate distances is proportional to the Casimir invariant
tr$\left\{ \lambda^a \cdot \lambda^a \right\}$ was actually identified as and important topic  by Del Debio et al. 
\cite{DelDebbio:1995gc,DelDebbio:1996xm}.
In what concerns SU(3), Deldar 
\cite{Deldar:1998ne,Deldar:1999vi,Deldar:2006hr} 
and Bali 
\cite{Bali:2000un} 
have also found a good agreement with Casimir scaling.  Note that the Casimir factors are computed analytically in group theory, see Table \ref{tab:casimir} for details. Bali, within an accuracy of 5 percent, found no violation to the Casimir scaling. 
To explain Casimir Scaling, Semay proposed that the cross section of the string is the same for all representations
\cite{Semay:2004br}.

Here we observe the colour flux tubes produced by a charge and an anti-charge of higher SU(3) representations. We extend the techniques developed by the previous authors both to study flux tubes and static potentials of higher SU(3) representations. Previously only the lowest 3 representation, and more recently the 8 representation were studied 
\cite{Cardoso:2009kz,Cardoso:2010kw}.

In this paper we present the results for the colour fields, the energy and Lagrangian densities as well as the respective Casimir scaling factors, using the energy density along the flux tube, for five different representations, computed in quenched SU(3) lattice QCD.
In section II, we introduce the lattice QCD formulation. We
briefly review the Wilson loops for this system, which was
used in \cite{Bali:2000un}, and show how we compute the colour fields as well as the Lagrangian and energy density distribution. In section III the numerical results are shown, including the chromoelectric field, chromomagnetic field, Lagrangian  and energy densities, the Casimir scaling factors and the flux tube widths for the various SU(3) representations. Finally, we present our conclusion in section IV.

\begin{figure}[t!]
\begin{center}
    \includegraphics[width=6.5cm]{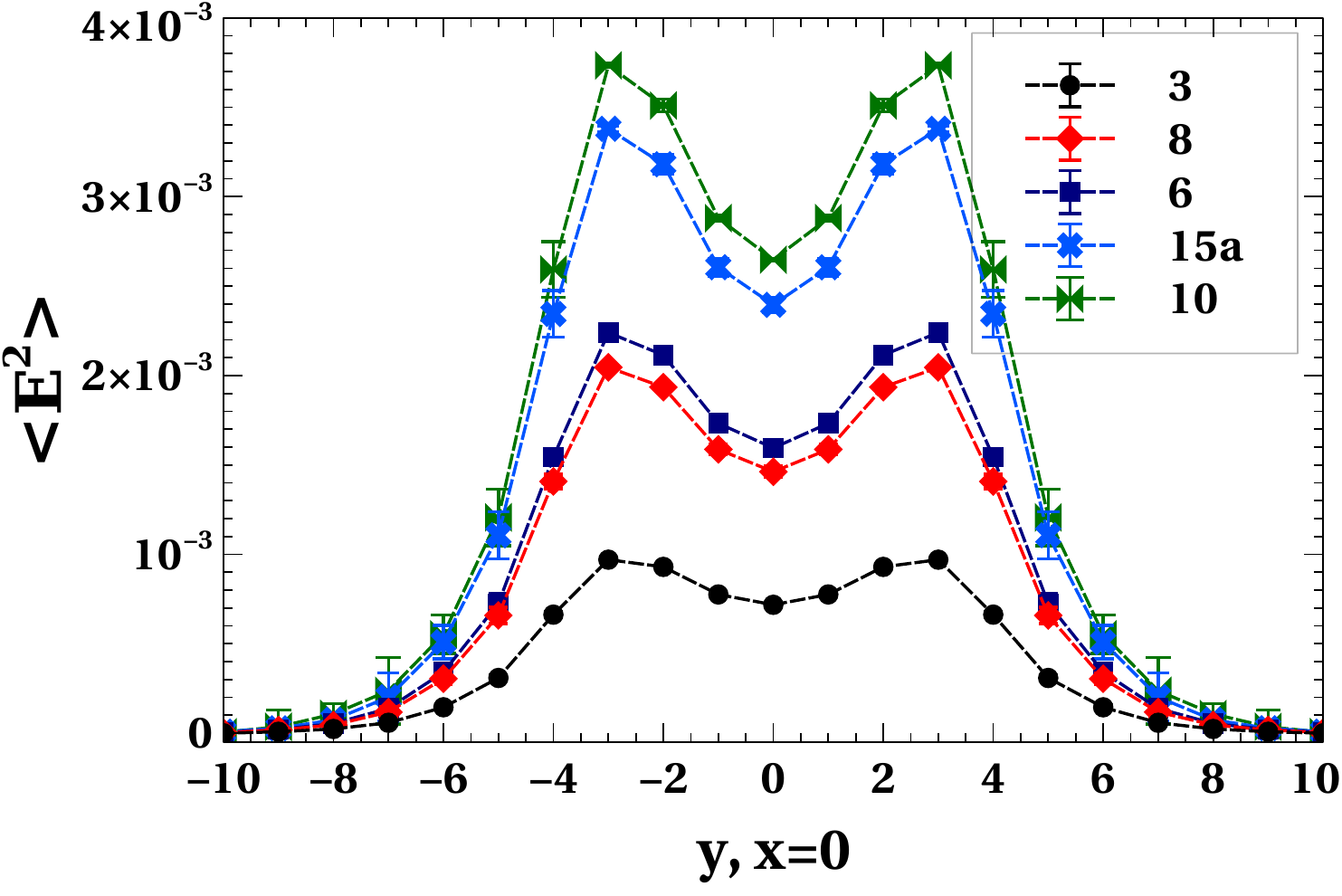}
    \includegraphics[width=6.5cm]{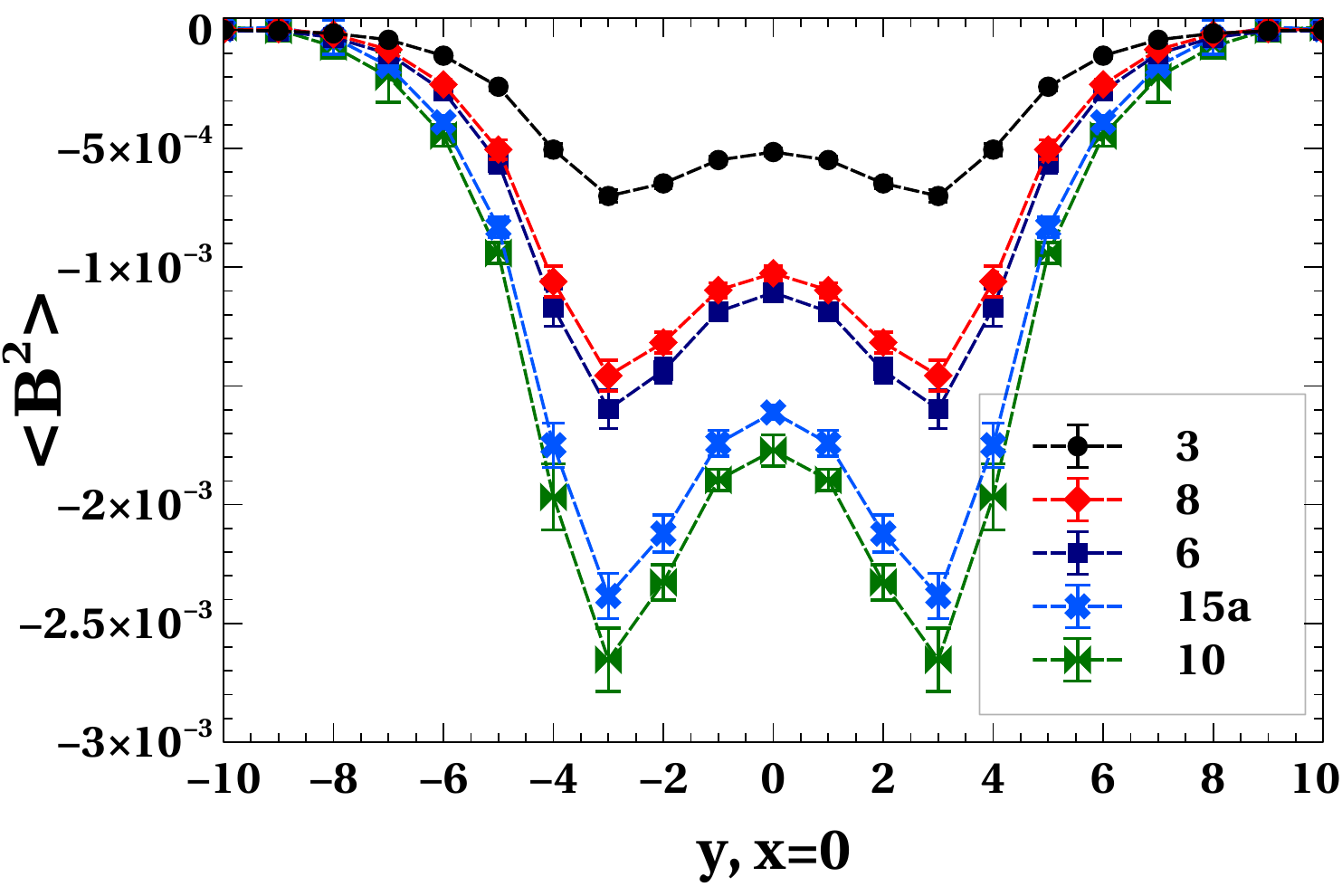}
    \includegraphics[width=6.5cm]{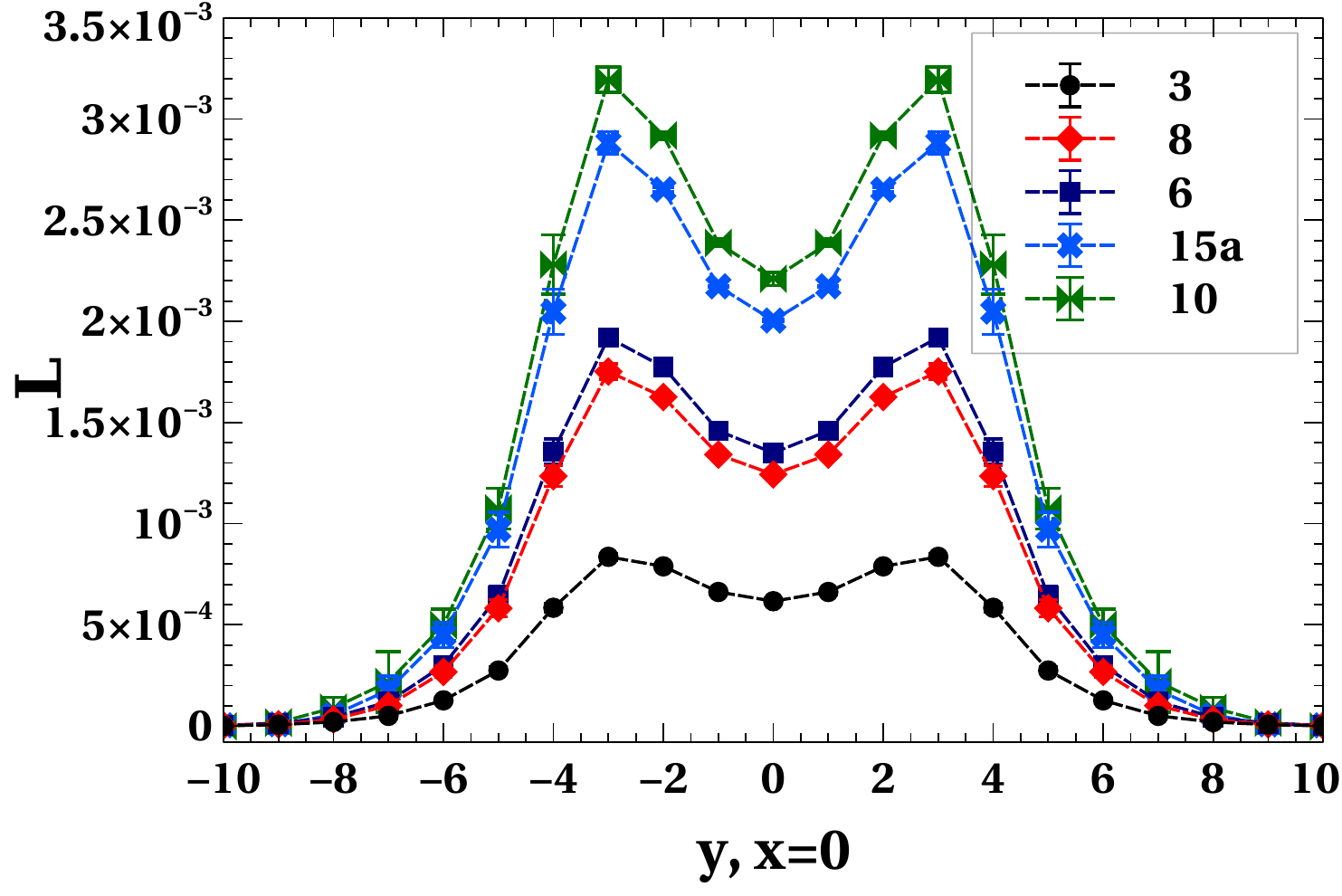}
    \includegraphics[width=6.5cm]{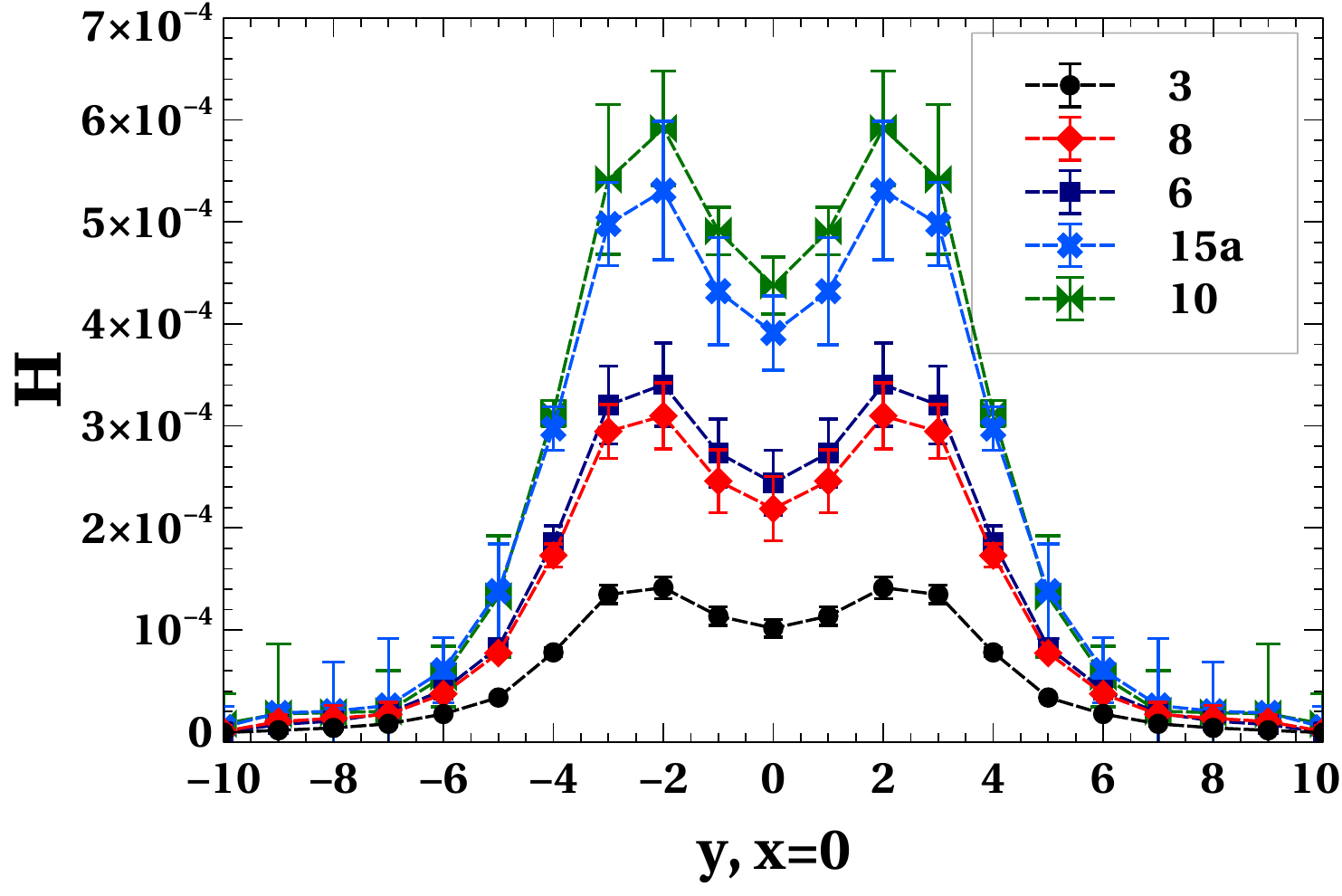}
   \caption{Chromoelectric field, chromomagnetic field, Lagrangian  and energy densities produced by the various SU(3) charges in the $y$ axis $(x=z=0)$. 
The charges are located at $y=\pm 4$.}
    \label{EBLE_ape_3hyp_x=0}
\end{center}
\end{figure}

\begin{figure}[!htb]
\begin{center}
    \includegraphics[width=6.5cm]{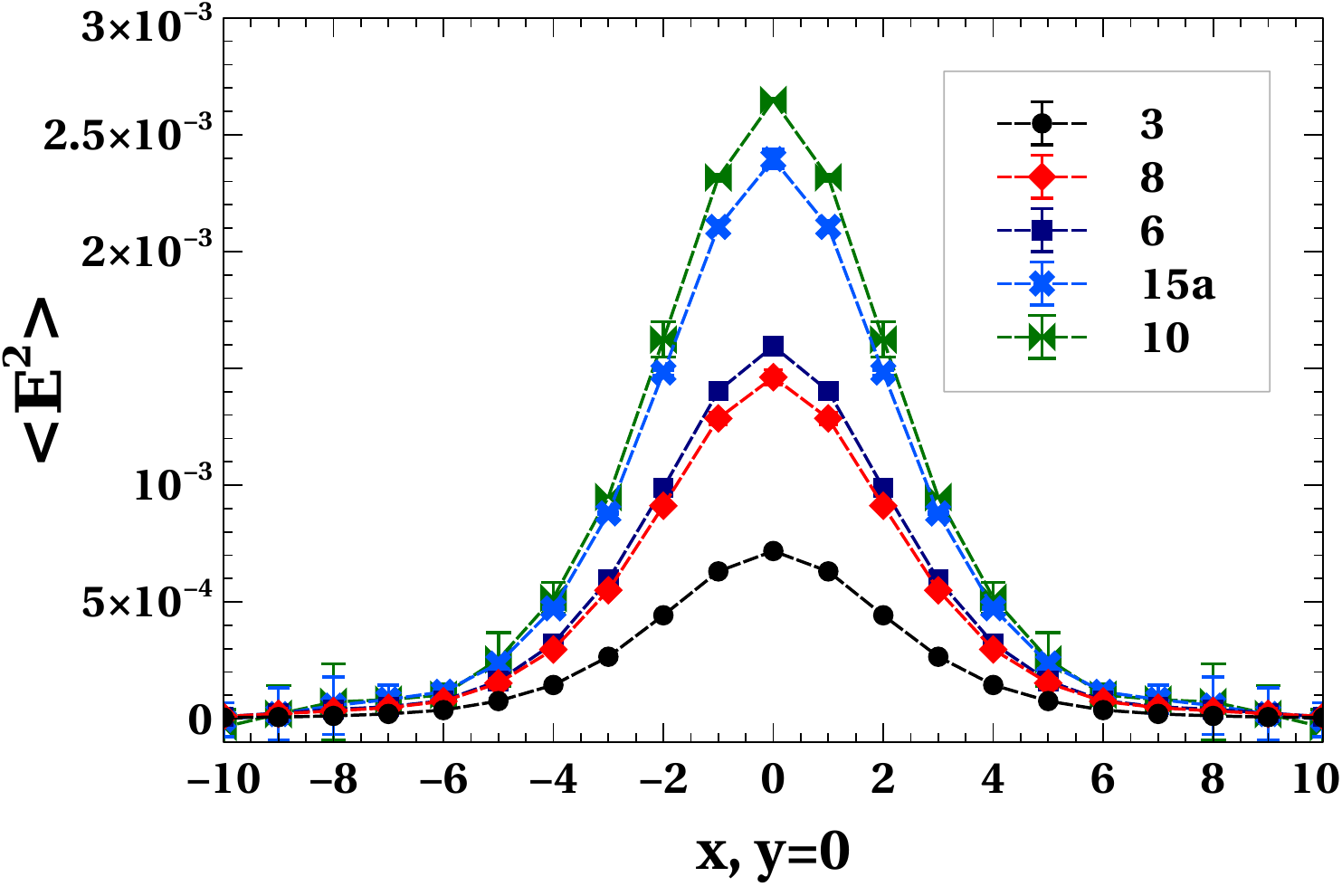}
    \includegraphics[width=6.5cm]{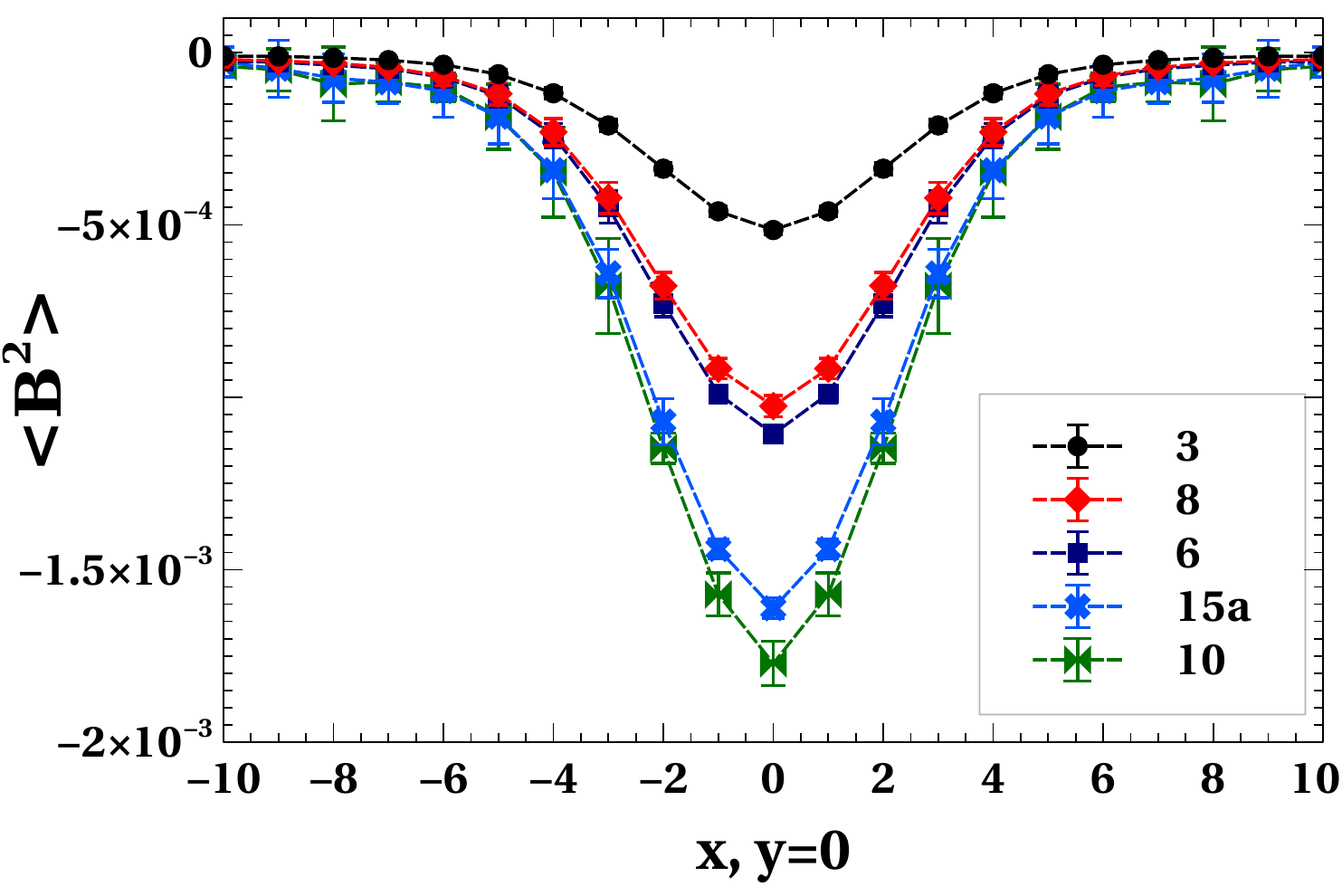}
    \includegraphics[width=6.5cm]{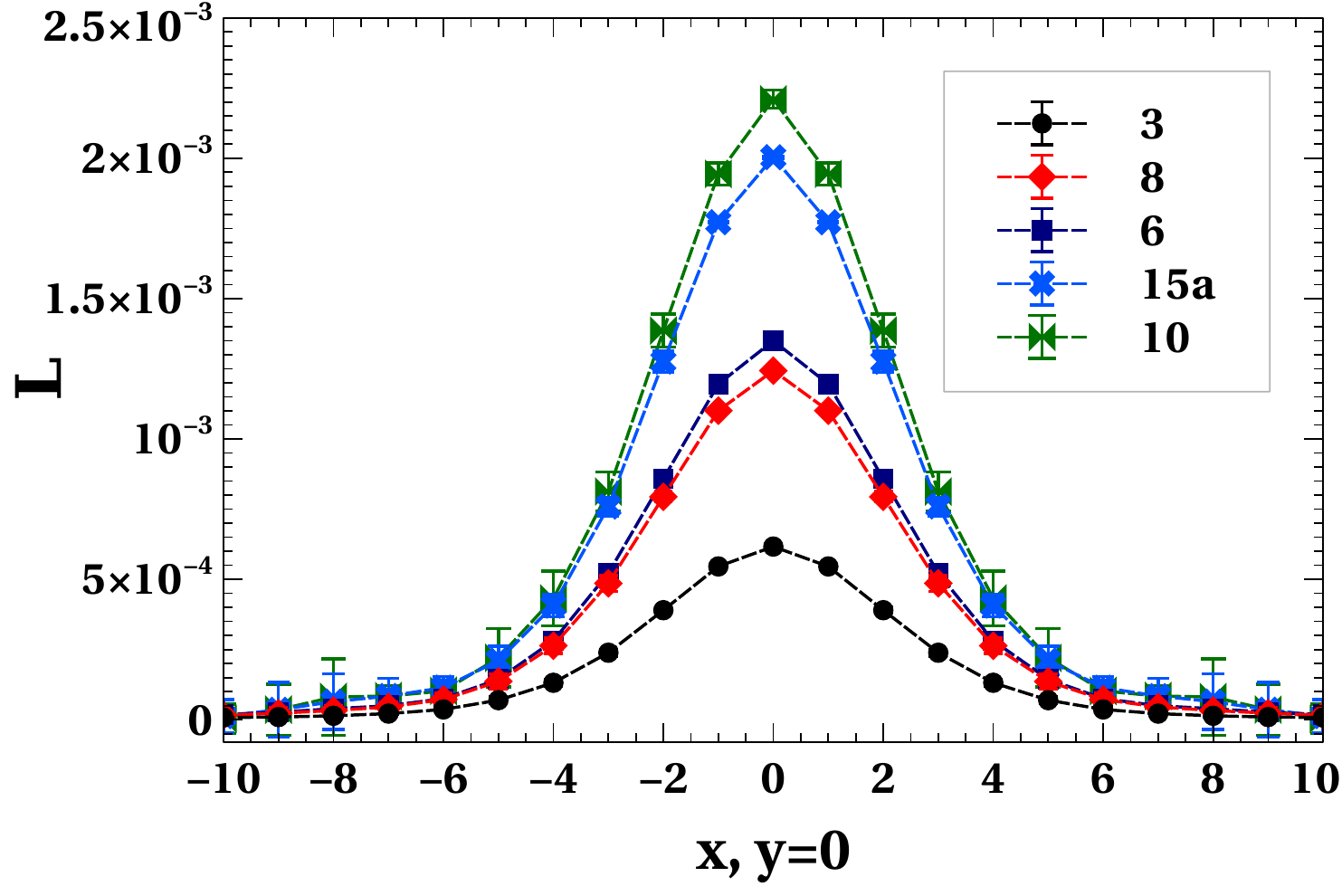}
    \includegraphics[width=6.5cm]{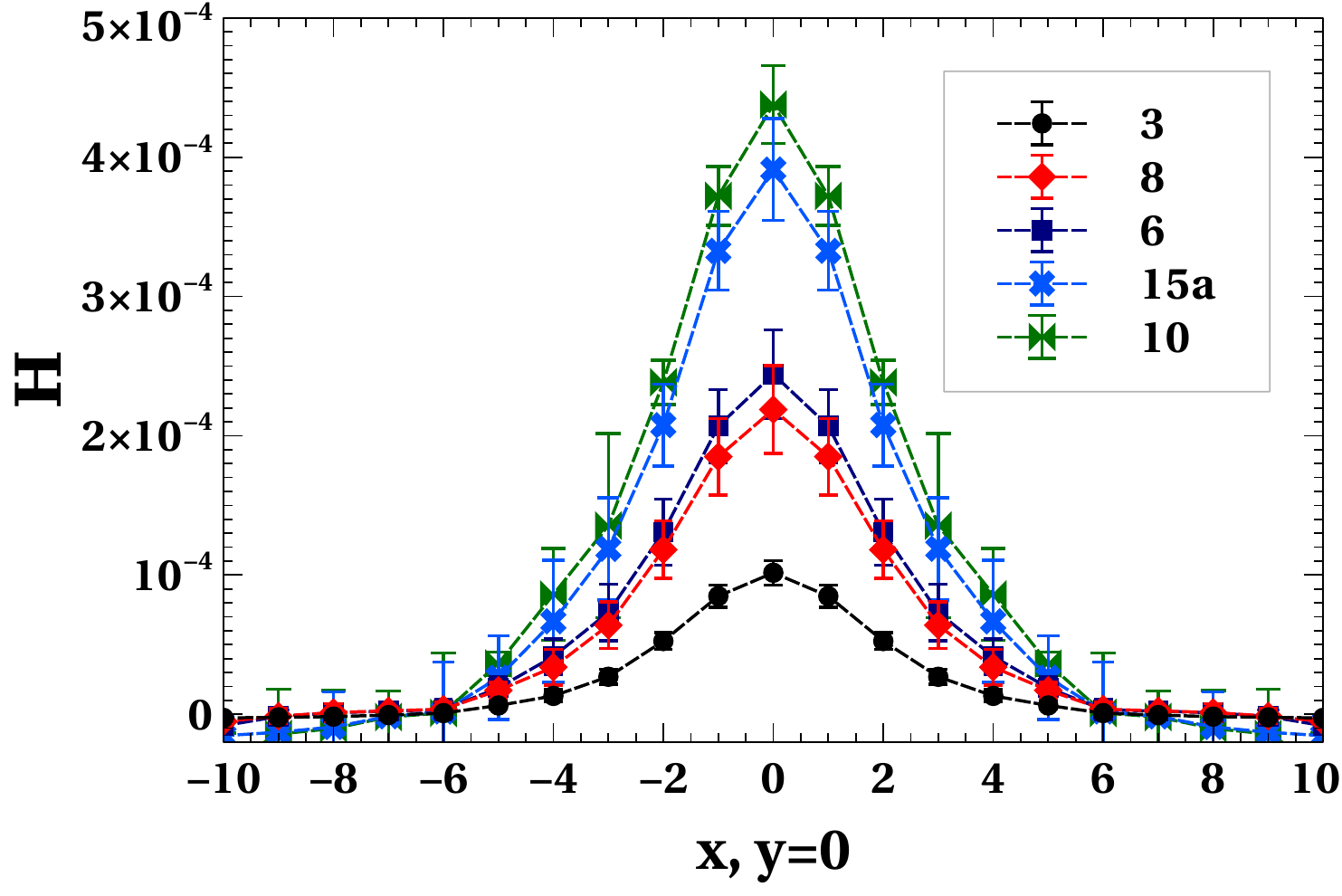}
    \caption{Chromoelectric field, chromomagnetic field, Lagrangian  and energy densities  produced by the various SU(3) charges along a radial distance in the metriatrix plane $y=0$.}
    \label{EBLE_ape_3hyp_y=0}
\end{center}
\end{figure}

\section{The Wilson Loops and Colour Fields}

To impose the position of the static charge and anti-charge in our Lattice, we utilize a Wilson loop.
The Wilson loop $W_D(\mbf r_1,\mbf r_2,T)$ creates a charge at position $\mbf r_1$ and an anti-charge  at $\mbf r_2$ separated by a spatial distance $R=| \mbf r_2  - \mbf r_1|$, which propagate for an Euclidian time $T$ maintaining their spatial positions $\mbf r_1$ and $ \mbf r_2$unchanged. To preserve gauge invariance, the wilson loop consists of a simple rectangular loop of size $R \times T$ composed of the trace of the product of links $U_D$ of the respective representation $D$ of SU(3).

Utilizing the algebra of the Gell-Mann generators of SU(3), the Wilson loops $W_D(\mbf r_1,\mbf r_2,T) $ for the various SU(3) representations 
\cite{Bali:2000un}, 
can be translated into products of the fundamental representation $D=3$ Wilson loops $W_3$, 
and fundamental links $U_3=U$. $U$ denotes a group element in
the fundamental representation of SU(3). The traces of $U_D$ in various representations, 
can easily be expressed in terms of traces of powers of $U$ 
as in Bali et al.
\cite{Bali:2000un}, 
\begin{eqnarray}
W_{3} & = & \Tr U\ ,
\non \\
W_{8} & = & \left|W_{3}\right|^{2}-1 \  ,
\non \\
W_{6} & = & \frac{1}{2}\left[\left(\Tr U\right)^{2}+\Tr U^{2}\right] \ ,
\non \\
W_{15a} & = & \Tr U\: W_{6}-\Tr U \ ,
\non \\
W_{10} & = & \frac{1}{6}\left[\left(\Tr U\right)^{3}+3\Tr U\Tr U^{2}+2\Tr U^{3}\right] \ .
 \end{eqnarray}

The chromoelectric and chromomagnetic fields on the lattice are given by the
Wilson loop and plaquette expectation values,
\bea
\label{fields} 
   \Braket{E^2_i(\mbf r)} &=& \Braket{P(\mbf r)_{0i}}-\frac{\Braket{W(r_1,r_2,T) \,P(\mbf r)_{0i}}}{\Braket{W(r_1,r_2,T)}} 
 \\ \non
    \Braket{B^2_i(\mbf r)} &=& \frac{\Braket{W(r_1,r_2,T)\,P(\mbf r)_{jk}}}{\Braket{W}(r_1,r_2,T)}-\Braket{P(\mbf r)_{jk}} \, ,
\eea
where the $jk$ indices of the plaquette complement the index $i$ of the magnetic field,
and where the plaquette at position $\mbf r=(x,y,z)$ is computed at $t=T/2$, 
\be
P_{\mu\nu}\left(\mbf r \right)=1 - \frac{1}{3} \ReC\,\Tr\left[ U_{\mu}(\mbf r) U_{\nu}(\mbf r+\mu) U_{\mu}^\dagger(\mbf r+\nu) U_{\nu}^\dagger(\mbf r) \right]\ .
\ee
Importantly, in SU(3) lattice QCD, since the Gel-Mann matrices have zero trace, the plaquette only produces the square of the components of the fields. This is consistent with the fact that the  colour field components are not gauge invariant (unlike the fields in electrodynamics). 

Nevertheless in SU(3) we can compute the Lagrangian ($\mathcal{L}$) and energy ($\mathcal{H}$) densities, since they are computed from the the square of the field components,
\bea
   \Braket{ \mathcal{H}(\mbf r) } &=& \frac{1}{2}\left( \Braket{\mbf E^2(\mbf r)} + \Braket{\mbf B^2(\mbf r)}\right)\ , \\
    \label{energy_density}
   \Braket{ \mathcal{L}(\mbf r) } &=& \frac{1}{2}\left( \Braket{\mbf E^2(\mbf r)} - \Braket{\mbf B^2(\mbf r)}\right)\ .
    \label{lagrangian_density}
\eea

To compute the static field expectation value, we plot the expectation value
    $ \Braket{E^2_i(\mbf r)} $ or   $\Braket{B^2_i(\mbf r)}$,  as a function of the temporal extent $T$ of
    the Wilson loop. At sufficiently large $T$, the groundstate corresponding to the
studied quantum numbers dominates, and the expectation value tends to a horizontal plateau.

In order to improve the signal to noise ratio of the Wilson loop, we use the APE smearing defined by
\begin{eqnarray}
    U_{\mu}\left(s\right) & \rightarrow & P_{SU(3)}\frac{1}{1+6w}\Big(U_{\mu}\left(s\right)\nonumber \\
    & & + w \sum_{\mu\neq\nu}U_{\nu}\left(s\right) U_{\mu}\left(s+\nu\right)U_{\nu}^{\dagger}\left(s+\mu\right)\Big)\ ,
\end{eqnarray}
with $w = 0.2$  
 and iterate this procedure 50 times in the spatial direction
\cite{Cardoso:2009kz}.
To achieve better accuracy in the flux tube, we apply three levels of hypercubic blocking (HYP) in the time direction, \cite{Hasenfratz:2001hp}, with
\begin{equation}
\alpha_1=0.75, \quad \alpha_2=0.6, \quad \alpha_3=0.3\ .
\end{equation}
Note that these two procedures are only applied to the Wilson Loop, not to the plaquette.
To compute the fields, we fit the horizontal plateaux obtained for each point $(x,y,z)$ 
determined by the plaquette position, but due to the azimuthal symmetry we only consider 
two coordinates for simplicity. 
For the positions $x$ (distance to the charge axis) and $y$ (distance along the charge axis) considered, except at the charge location, we find in the range of $T\in [3,12]$ in lattice units, horizontal plateaux with a $\chi^2$ /dof $\in [0.3,2.0] $.  
We find these horizontal  plateaux for representations up to $D=10$. 
For higher representations then the statistical noise increases significantly
already at $T$ below 12, and thus we decide not to further study representations
higher than $D=10$.
We finally compute the error bars of the fields with the jackknife method.

\section{Results}

\begin{figure}[t!]
\begin{center}
    \includegraphics[width=8.5cm]{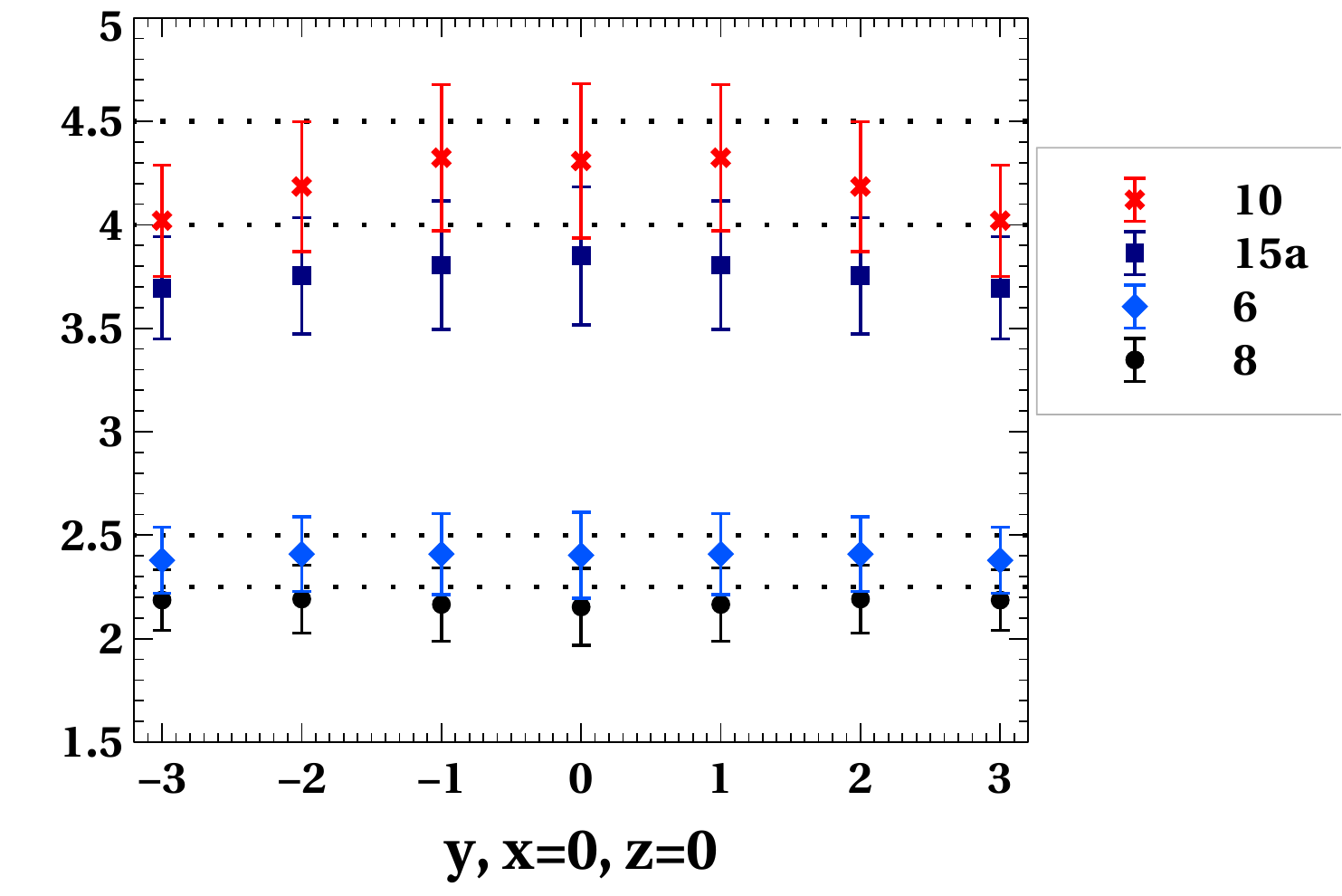}
    \caption{Casimir scaling of the energy density for the various SU(3) representations in the charge axis.}
    \label{Casimir_ape_3hyp_x=0}
\end{center}
\end{figure}

\begin{figure}[t]
\begin{center}
    \includegraphics[width=8.5cm]{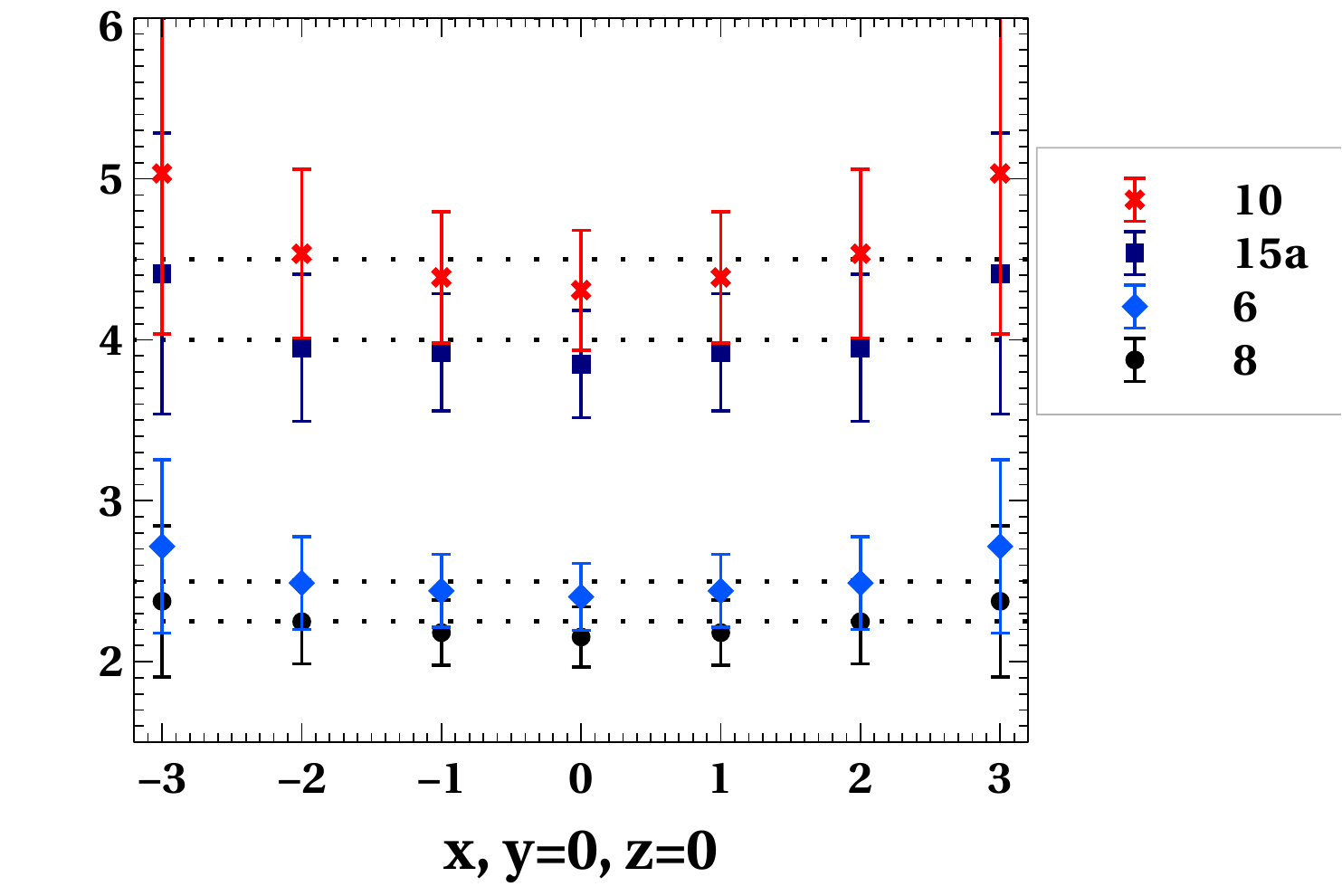}
    \caption{Casimir scaling of the energy density for the various SU(3) representations in the mediator plane.}
    \label{Casimir_ape_3hyp_y=0}
\end{center}
\end{figure}

Here we present the results of our simulations with 381 quenched configurations in a $32^3 \times 64$ lattice at $\beta = 6.2$.
We generate our configurations in NVIDIA GPUs of the FERMI series (480, 580 and Tesla 2070) with a SU(3) CUDA code
upgraded from our SU(2) combination of Cabibbo-Marinari
pseudoheatbath and over-relaxation algorithm \cite{Cardoso:2010di,ptqcd}.
Our SU(3) updates involve three SU(2) subgroups, we work with 9 complex numbers,
and we reunitarize the matrix.
We have two options to save the configurations, either in a structure of arrays where
each array lists a given complex component for all the lattice sites, or
in an array of structures where each structure is a SU(3) matrix. 

The results are presented in lattice spacing units of $a$, with $a=0.07261(85)\,$ fm or $a^{-1}=2718\,\pm\, 32$ MeV.
We locate our colour charge and anti-charge  at $\mbf r_1=(0,-4,0)$ and $\mbf r_2=(0,4,0)$.

\subsection{Colour Fields}

In Figure \ref{fig:3D_Sim} we depict the 3D plots of the flux tubes produced by the various SU(3) charges. 

To arrive at a more quantitative picture of the flux tubes, we cut the flux tubes in their symmetry axis. We depict in Figure \ref{EBLE_ape_3hyp_x=0} the profiles of the different components of the flux tube, in a plane parallel to the quark-antiquark axis. It is is well known that the APE smearing and HYP blocking affect the fields in the charge neighbourhood
\cite{Okiharu:2004tg}, 
decreasing the respective peaks and shifting them slightly from the $\mbf r_1$ and $\mbf r_2$, but do not affect the aim of our study which is the study of the flux tubes.

We also show in Figure \ref{EBLE_ape_3hyp_y=0} the profiles of the different components of the flux tube, in the mediator plane, perpendicular to the quark-antiquark axis.

The profiles of the different representations are similar in shape, only changing in magnitude,
as suggested by Semay to understand Casimir Scaling
\cite{Semay:2004br}.

\subsection{Casimir Scaling}

We further compare the different SU(3) flux tubes with the Casimir scaling.
 
In the Figs. \ref{Casimir_ape_3hyp_x=0} and \ref{Casimir_ape_3hyp_y=0}, we show the results for the ratio between the energy density of every representation over the energy density of the fundamental triplet representation. We compare these ratios with the ratio of the Casimir scaling hypothesis. is correct. Indeed, at least for the first five lower representations of SU(3), and in the centre of the flux tube where most of the colour field intensity is concentrated, our results are in agreement with a flux tube Casimir scaling, within error bars.

\subsection{Flux tube width}

Moreover, we determine the width of the various SU(3) flux tubes, computing the Radius Mean Square in the mediator plane $y=0$,  utilizing the Lagrangian density as a probability density,
\be
RMS = \sqrt{ \sum_x  x^3 \, {\cal L}(x) \over \sum _x x \, {\cal L}(x) } \ ,
\ee
evaluated only in the positive $x$ direction.
In Table \ref{tab:RMS} we show the respective $RMS$ obtained for the various SU(3) representations.
We find a similar Radius Mean Square for the various SU(3) flux tubes we study,
of the order of 0.35 fm. This is consistent with the conjecture of Semay 
\cite{Semay:2004br}
for a constant cross section of the various flux tubes.

\section{Conclusions}

\begin{table}[!t]
\begin{center}
\begin{tabular}{ccc}
\hline
	$D$ & &$RMS$ [fm] \\
\hline
\hline
3 &  &$ 0.351 \pm   0.023 $ \\
8 &  &$ 0.343  \pm      0.028 $ \\
6 &  &$ 0.392  \pm     0.029 $ \\
$15a$ & & $ 0.312 \pm   0.054 $ \\
10 &  & $ 0.395  \pm     0.123 $ \\
\hline
\end{tabular}
\end{center}
\caption{
The different Radius Mean Squares obtained for each of the SU(3) representations we study here.}
\label{tab:RMS}
\end{table}

We study in quenched Lattice QCD the flux tubes produced by static SU(3) colour charges.
This is relevant to understand the mechanisms for the confinement of colour SU(3) charges, and also to find a solid foundation for the very strong and thin flux tubes utilized in models of Heavy Ion Collisions. 

For the first five representations 3, 8, 6,  15$a$ 	and 	10, up to Casimir scaling factors of
4.5, our Wilson loop plateaux have acceptable $\chi^2$ /dof $\in [0.3,2.0] $ and we depict the 
different colour field components.

Our computations show that the energy density of the flux tubes produced by the different static SU(3) sources compatible with Casimir scaling. The energy density, within error bars, is proportional to the eigenvalue of the quadratic Casimir operator of that representation.

All the flux tubes of the different representations show a similar radius mean squares circa 0.35 fm.
This is twice as large, but still of  the same order, of the width of the thin flux proposed to exist at the onset of Heavy Ion Collisions.  
It remains possible that the so-called perturbative flux tubes of Heavy Ion Collisions 
evolve to the non-perturbative flux tubes observed here.

\section*{Acknowledgments}
We thank Edmond Iancu, Gunnar Bali, Jeff Greensite, Magdalena Malek and Orlando Oliveira  
for useful discussions.

This work was partly funded by the FCT pluri-annual contract, 
PTDC/FIS/100968/2008,  and the FCT annual contracts 
CERN/FP/109327/2009 and 
CERN/FP/116383/2010.
Nuno Cardoso is supported by the FCT fellowship SFRH/BD/44416/2008.

\bibliographystyle{model1-num-names}
\bibliography{repsu3}

\end{document}